\documentclass[a4paper,11pt]{article}
\usepackage{pos}

\title{Exploring GRB Afterglows in the TeV Era: New Diagnostics of Particle Acceleration}

\author*[a]{L. Foffano}
\author[a]{M. Tavani}

\affiliation[a]{INAF -- IAPS, via del Fosso del Cavaliere 100, I-00133 Roma (Italy)}

\emailAdd{luca.foffano@inaf.it}

\abstract{
The TeV gamma-ray band is essential for probing the most extreme particle acceleration processes in the Universe. The recent detections of gamma-ray bursts (GRBs) at these energies offer an incredible opportunity to investigate the origins of such transient events in an unprecedented way. In this presentation, we analyze the afterglows of these GRBs by modeling their synchrotron and inverse Compton emission within an optimized relativistic fireball framework. By comparing observational data with theoretical predictions, we constrain key model parameters and track their temporal evolution. The comparison of different TeV-detected GRBs reveals an intriguing variety among them, potentially reflecting differences in the particle acceleration processes that have to be very fast and able to accelerate to large energies. We discuss how late-time afterglow observations of X-ray and GeV-TeV emissions are crucial for providing diagnostics into the physics of GRBs. At this scope, we also present the most updated results of the AGILE telescope, which support our interpretation. Finally, we highlight theoretical predictions for future TeV observations and their implications for understanding these extreme cosmic explosions.
}

\ConferenceLogo{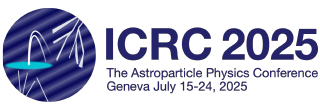}

\FullConference{39th International Cosmic Ray Conference (ICRC2025)\\
 15–24 July 2025\\
Geneva, Switzerland\\}

\begin{document}
\maketitle

\section{Introduction} 
\noindent
Gamma-ray bursts (GRBs) are extremely energetic explosions that push particle acceleration mechanisms to the limits of known physics. 
Observationally, they are subdivided on the basis of their duration: while long-duration GRBs (lasting $>$2~s) are commonly associated with the core collapse of massive stars, short-duration GRBs (lasting $<$2 s) are often linked to neutron stars merging events. 

GRBs produce collimated jets of plasma where particles are accelerated at ultra-relativistic speed, thanks to extreme particle acceleration mechanisms \citep[see e.g. a recent review by][]{2018IJMPD..2742003N}.
After an intense prompt emission connected to the dynamical activity of the implosion, the shock waves of the GRB expand into the external medium. Particles are accelerated by the shocks, producing broad-band emission across the whole electromagnetic spectrum.

\section{TeV gamma~rays from gamma-ray bursts: a new era of investigation}
\noindent
GRB afterglows have been extensively studied for several years, primarily in key energy bands such as optical and X-rays.  However, a new era into the investigation of these extreme phenomena has opened with the detection of very-high-energy (VHE, >100 GeV) gamma rays from the afterglow of several GRBs in recent years.
The confirmation of strong highly energetic radiation emitted by these cosmic explosions supports an interpretation in terms of inverse Compton (IC) reprocessing of synchrotron photons \citep{2001ApJ...559..110Z}, highlighting the role of the Synchrotron Self-Compton (SSC) model to describe the emitted TeV photons \citep{sari_esin_2001}.

The dynamical evolution of these GRBs has, for several decades, been interpreted within the framework of the \textit{``relativistic fireball model''} \citep[e.g.,][]{Rees_1992_relativistic_fireball, 1997ApJ...476..232M, 1998ApJ...499..301M, Piran_review_GRB_1998, Chiang_dermer_1999, Panaitescu_2000A}.
Even though this theoretical model has been broadly confirmed by the latest TeV gamma-ray observations, the data also suggest a more complex overall evolutionary scenario that remains a topic of ongoing debate.
It turned out that the systematic observations of GRBs across a wide range of energy bands, extending to very early times after the burst onset, reveal global observational properties that challenge the traditional GRB paradigm. These observations show flux evolutions and spectral behaviors that are not fully compatible with the expected global evolution predicted by standard models. Understanding these differences -- which are deeply connected with the extreme underlying particle acceleration processes -- is crucial to develop a comprehensive theoretical framework for these events.

\section{New diagnostics to explore the particle acceleration mechanisms underlying gamma-ray bursts}
\noindent
In this work \citep[][]{Foffano_2025}, we explore the information provided by the most energetic emission produced by these extreme transient phenomena, following our previous activity reported in \citet{foffano2024_GRB221009A} and \citet{Tavani_2023}. Specifically, we investigate all GRBs currently detected at TeV energies, aiming at comparing their VHE emission. By applying an optimized version of the relativistic fireball model to a selected sample of broad-band observational data, we take into consideration the information emerging from such a comparison. Our analysis adopts a global view of the temporal evolution of the flux and of the broad-band spectral emission,  aiming to uncover new diagnostics of the underlying particle acceleration mechanisms in these extreme transient events.

\section{Results and Conclusions}
\noindent
The outcome of our global analysis over the known TeV GRB afterglows supports the importance of late-time afterglow observations of X-ray and GeV-TeV emissions, which are crucial for providing diagnostics into the physics of GRBs. At this scope, we also present the most updated results of the AGILE telescope on a set of recent GRB events, which support our interpretation.

Our results show that the TeV afterglow lightcurves may possibly show a steepening of the lightcurves at moderately late times. These decays can be caused by different effects, either jet breaks (in the simple top-hat case or also in the structured jet scenario) or intervening spectral effects.

Our analysis concludes that the advent of systematic TeV gamma-ray follow-up observations of GRB afterglows will provide important information to investigate further the underlying particle acceleration mechanisms. 
Furthermore, we support the adoption of both early and medium-term follow-up observations: 
\begin{itemize}
    \item early observations -- up to a few hours from the trigger time --  can investigate the transition from the prompt phase to the earliest afterglow phase;
    \item later observations -- extending up to a few days after trigger -- offer the opportunity to explore important features of the physics underlying the extreme particle acceleration in long GRBs.  In particular, the {TeV lightcurve steepening} can be detectable with gamma-ray observations even after one or two days after trigger. 
\end{itemize}
It is worth noting that these late-time TeV observations would also support the observational limits of Cherenkov instruments, especially concerning their reduced sensitivity for short observations and the dependence on the sky quality during observations. 
Late-time observations, providing longer exposure times compared to early-time ones, lead to improved sensitivity and compensate for potential non-optimal atmospheric conditions at the trigger time of the events. In this context, our results underline the relevance of late-time TeV observations on the physical point of view, especially in light of the forthcoming generation of Cherenkov telescopes, which will benefit of much improved sensitivity in the TeV gamma-ray band.


\bibliographystyle{unsrtnat}
\bibliography{biblio}{}

\end{document}